# Structured Optical Materials Controlled by Light

Sara Nocentini[1,*], Daniele Martella[1,2], Camilla Parmeggiani[1,2,3], Simone Zanotto[1,*,+],
Diederik S. Wiersma[1,4,*]

[1] European Laboratory for Non Linear Spectroscopy, via N. Carrara 1, Sesto Fiorentino, Firenze 50019, Italy
[2] Chemistry Department "Ugo Schiff", University of Florence, Via della Lastruccia 3, Sesto Fiorentino, Firenze 50019, Italy
[3] CNR-INO Sesto Fiorentino, Via Madonna del Piano 10, Firenze 50019, Italy
[4] Physics Department, University of Florence, via G. Sansone 1, Sesto Fiorentino, Firenze 50019, Italy
[+] Present address: NEST, Istituto Nanoscienze - CNR and Scuola Normale Superiore, Piazza S. Silvestro 12, Pisa 56127, Italy
[*] E-mail: nocentini@lens.unifi.it, simone.zanotto@nano.cnr.it, wiersma@lens.unifi.it

Materials of which the optical response is determined by their structure are of much interest both for their fundamental properties and applications. Examples range from simple gratings to photonic crystals. Obtaining control over the optical properties is of crucial importance in this context, and it is often attempted by electro-optical effect or by using magnetic fields. In this paper, we introduce the use of light to switch and tune the optical response of a structured material, exploiting a physical deformation induced by light itself. In this new strategy, light drives an elastic reshaping, which leads to different spectral properties and hence to a change in the optical response. This is made possible by the use of liquid crystalline networks structured by Direct Laser Writing. As a proof of concept, a grating structure with sub-millisecond time-response is demonstrated for optical beam steering exploiting an optically induced reversible shape-change. Experimental observations are combined with finite-element modeling to understand the actuation process dynamics and to obtain information on how to tune the time and the power response of this technology. This optical beam steerer serves as an example for achieving full optical control of light in broad range of structured optical materials.

## 1. Introduction

Material structuration at the nano/micrometric scale over different architectures impresses peculiar optical properties as result of light scattering and wave interference. Natural and synthetic materials with a periodic modulation in their refractive index diffract light of wavelengths commensurate to their periodicity and, depending on their arrangement and material, they constitute photonic structures of different complexity, ranging from one dimensional gratings (Bragg grating) up to 3D photonic crystals. In all these cases, light-matter interaction leads to the characteristic optical features, as for example, reflection and transmission profiles of diffractive gratings and even more interestingly the propagation inhibition of certain frequencies (photonic band gap) in photonic crystals. Surprisingly, numerous examples of how nanometric element disposition affects the optical properties are provided also by nature. In fact, structural coloration,[1] a nanometric arrangement of dielectric geometric units on the skin or surface of animals, is an efficient strategy adopted by many living organisms to create vivid and bright structural colors to adapt to the surrounding environment, to send warning messages or mislead their natural enemies through camouflage.[2-4] Similarly to what happens in nature, which provides actual reconfigurable mechanisms,[5] also in man-made photonic applications, it is very important to have control over the spectral response of a device or material. Current strategies include the use of the electro-optic effect,[6] temperature sensitivity,[7] and carrier injection.[8] Another method relies on the birefringent behavior of liquid crystals,[9] which can be infiltrated in photonic structures allowing to control the refractive index through temperature,[10,11] or electric[12] or magnetic[13] fields. Materials with structural color can be effectively - but invasively and slowly - tuned by deforming them, for instance by applying mechanical pressure or stress.[14-18] This is the only proposed tuning method over the spectral response that acts on the unit cell variation instead that on refractive index control. Here we wish to explore a different route to achieve tuning and switching of photonic materials, using the light itself as means of controlling the mechanical state of a periodic material. By creating a structured material out of photo-responsive polymer, we show that it is possible to induce mechanical deformation – and hence a huge change in the optical response of the material – by simply shining light on it. De facto, this constitutes an enormous non-linear optical response, with sub-millisecond time-response, using mechanical deformation as intermediate step. The deformation, and hence the non-linear response, takes place only in the illuminated region and therefore can be considered local to a high degree. The photo-responsive polymers that we use are light sensitive liquid crystal networks (LCN), or elastomers, which have the capacity to strongly deform in a reproducible way.[19] It has been demonstrated that LCN can be patterned[20] to act as actuators both in photonic applications[21] and in robotics.[22,23] In this work, we exploit their potential as polymeric matrix for the realization of tunable photonic structures. The idea is to create photonic materials out of light sensitive polymer such that the structure deforms (locally or entirely) in a controlled way upon illumination, thereby changing the photonic properties. As first demonstration of this concept, we created an optically controlled two dimensional diffraction grating. Despite its simple structure, such device has an important application, namely that of active beam steering. Beam steering is relevant in several sectors of photonics, including microscopy,[24] displays,[25] object detection,[26] and optical tele- and data communication.[27,28] The existing technology involves devices like MEMS mirrors[29-30] or, for instance, phase gratings,[31-33] and is based on electrical,[34] electromechanical[35] or acousto-optic[36] effects. Nonetheless, optically tunable photonic components have been demonstrated for structures made by GST ($Ge_3Sb_2Te_6$) and $VO_2$ but their very complex, subwavelength-scale structuration (phase array devices)[37-43] requires extremely precise nanofabrication. Our beam steerer, on the other hand, provides fully optical control and simple integration at the microscale. Moreover, the variation of the geometrical pattern within this application that depends on structural periodicity enlarges the tuning range. To learn something about the advantages and disadvantages, and the future potential of the new strategy proposed in this paper, we investigate the temporal and power response of this device in detail, changing geometrical and environmental parameters. The study involves both the realization of materials, experimental characterization of the devices, and theoretical support via numerical calculations. We observed a strong steering effect with response times in the sub-millisecond range. Excitation powers are in the range of milliwatts. Both response time and sensitivity depend on the shape and thermal conductivity of the environment. We found that the soft polymeric matrix provides an efficient energy transfer mechanism from the driving light beam to the mechanical and optical response of the structure. The results of this study show that micro-structured photo-sensitive polymers can provide an interesting new platform in the research and realization of fully optically controlled photonic devices.

**2. Material and lithographic technique**

Advances in chemical engineering of polymeric materials made them appealing in view of photonic applications. In fact, due to their low cost, easy fabrication and good optical properties, they constitute an ideal platform to further incorporate more complex materials whose hallmarks match the desired physical and chemical properties for the system under study. In fact, polymers with different characteristics (conductive,[44] biocompatible,[45,19] deformable,[46-47] hydrophilic/hydrophobic[48]) can be patterned even with a nanometric resolution in order to create functionalized responsive photonic microdevices. Here we employed a photoresponsive elastic matrix made of liquid crystalline networks.

**2.1 Liquid crystalline networks**

Liquid crystal networks (LCN) are smart materials that can introduce new functions in photonic applications. They combine the well-known liquid crystal properties with the elasticity of rubber polymers. The typical liquid crystal (LC) molecular alignment gives rise to a birefringent material with a higher refractive index along the director (the preferential direction along which the LC molecules tend to align) and determines also the deformation nature. In fact, depending on the molecular alignments various movement such as contraction,[49] bending[50] or rotation[51] can be achieved. The material reshaping follows the molecular rearrangement while excited: it starts losing its typical anisotropic long range order and once the elastomer overcomes the phase transition it reaches a more isotropic molecular organization. Therefore, liquid crystalline transitions affect both the optical properties and the geometrical ones, determining a refractive index variation and an elastic deformation, respectively. This is a unique case among the materials for photonics applications where both effects contribute to the photonic reconfiguration of the structure.

The LCN mixture composition employed here has already been described.[52] It is mainly constituted by two different LC mesogenic monomers, M1 and CL1. The latter molecule, bearing two acrylate groups, behaves also as a cross-linker giving reversible shape-changing properties to the material. Adjusting the ratio between these mesogens allows to improve the resolution for sub-micrometric objects fabricated with the DLW[20] and to modify the dynamics of the LCN deformation under light activation.[53] For this application, the required rigidity to have a free-standing grating with a well-defined pitch, was obtained by a mixture with 30% of mol/mol of cross-linking agent. The mixture also contains an initiator (IN1,1% mol/mol) which triggers the photo-polymerization and an azobenzene monomer (D1, 1% mol/mol), whose absorption peak is in the green region of the visible spectrum, which works mostly as a nanoscale heater making this LCN mixture a photoresponsive material.[22]

To control the molecular alignment, and therefore the final structure deformation, glass cells with a chemical and mechanical treatment were employed. A homogeneous uniaxial alignment was selected in order to have an in-plane asymmetric deformation. The monomeric mixture is optimized to reach the nematic phase at room temperature and to maintain it for several hours without crystallization, thus the so prepared LC cell can be directly used for Direct Laser Writing. The lithographic process leads to the fabrication of polymeric structures with retention of the nematic alignment in the final, non-liquid material.

## 2.2 Three dimensional nanometric patterning with laser lithography

Patterning at the sub-micrometric scale was performed with a two-photon photo-polymerization process (Direct Laser Writing, DLW).[53] With this system, a three dimensional soft photonic structure with features as small as 300 nm can be realized.[20] The 3D polymerization allows not only to create the planar grating structure (**Figure 1**a) but also to suspend it on pillars that have different functions. First, they separate the LCN structure from the substrate increasing the maximum allowed deformation while preventing the structure from binding to the glass. Second, they influence the thermal dynamics of the structure, which is in turn connected with the actuation time (see details in the following). Deviations from the ideal periodicity are caused by the monomer diffusion (in few seconds) through the already cross-linked structures during the polymerization process, giving rise to a line warping effect. Further chemical engineering of the monomer mixture would reduce this effect, which however affects only marginally the effectiveness of the tunable grating operation. After the fabrication (Figure 1a), the unpolymerized mixture is removed by immersing the glass cell in 2-propanol at 50 °C leaving the micrometric polymeric grating attached to the glass through the pedestals (Figure 1b-c). The grating pitch is 1.5 μm. The same periodic structure has been realized on pedestals of different heights to study the deformation dynamics as a function on the grating-glass distance.

## 3. Discussion and results
## 3.1. Evaluation of Diffraction Tunability

The operation principle of the remotely tunable grating beam steerer is illustrated in **Figure 2**a: when a control beam (green laser) excites the elastic structure, the LCN molecular arrangement switches from its ordered nematic arrangement to a more disordered phase. Thanks to elasticity guaranteed by the crosslinker, the whole structure deforms reversibly and anisotropically, accordingly to the LCN alignment director determined during the cell fabrication. The grating pitch varies by almost equal and opposite amounts in the two orthogonal directions, and as a consequence of the grating reshaping, a red probe beam impinging normally to its surface, it is then diffracted at different angles. The rest and excited configuration of the diffraction profile are reported in Figure 2b and c. In this experiment, a micro-beam steering constituted by a 2D grating with a pitch of 1.5 µm, hold by a matrix of 3x3 pedestals (5 µm long) was employed. As reported in Figure 2d and e, by adjusting the activation power, it is possible to control the extent of the deformation which depends from the temperature increase induced by the irradiation[22] in a quasi-complete reversible way. Thus, the beam deviation angle depends on the power applied through the control beam opening to possibly address adjacent positions tuning the incident power value. For the horizontal diffracted orders, we recorded an increase of the deviation angle that corresponds to the stretching of the LCN structure perpendicularly to the LC molecules director. At the maximum applied green laser power of 19.5 mW we observed a deviation, of the ±1H horizontal diffracted orders of 20-22%. The difference between the two values of the positive (+1H) and negative order (-1H) has to be attributed, likely, to the imperfections of the analyzed device. For the same reason the +1H beam does not return exactly back to the initial position after performing the complete power cycle.

Consistently with the anisotropic deformation of the grating (i.e., contraction along the LCN director and elongation along the orthogonal direction), the ±1V vertical diffracted beams instead get closer to the zero-order transmitted beam (Figure 2e). Here the deviation of the diffracted beams is slightly less than 20%, revealing that the grating does not contract/expand by the exact same amount along the two directions. Nonetheless, the effect is very clear and can be employed to effectively steer multiple beams.

**3.2 Submillisecond actuation dynamics**

As for every actuator, an essential aspect of our device is the time-response. To characterize the dynamical behavior of the beam steering we modified the setup employed for the above described experiment: the diffraction pattern is not formed any more on a screen, rather, it is directly imaged by a Fourier lens on the sensing element of a fast CCD. The control beam is chopped by a rotating wheel at 100 Hz and directly monitored by extracting a fraction. We reported here the dynamic characterization of a LCN grating with the same in-plane periodicity of 1.5 µm, 5 µm tall pedestals arranged in a 3x3 matrix. **Figure 3**a reports the temporal trace of the beam steering, intended as the angular shift due to the excitation divided by the diffraction angle of the unexcited grating. The +1H diffracted beam is considered, although equivalent traces are observed from the other beams. A regular periodic behavior is observed, with almost exponentially rising and falling fronts (Figure 3b - c). The actuation times (i.e., the times $t_{contr/relax}$ entering the exponential factor $\exp(-t/t_{contr/relax})$) observed for various control power levels are reported in Figure 3d, along with the steady-state beam deflection. The increase of the control power determines a bigger deformation as the local temperature reaches a value closer to the phase transition temperature. At the same time, this stored energy inside the structure affects also the time-response, speeding up the beam steering process. A slight difference in the power levels with respect to the observations reported in the previous section are due to differences in the focusing optics employed at this stage. Notably, it turns out that the LCN grating is capable of submillisecond operation: the rise times are slightly below the millisecond value, and the decay times are almost the half, below the 0.5 ms, or even 0.36 ms as in the case of 2.25 mW of excitation power. This result exceeds by several orders of magnitude the typical values reported for light-driven polymeric tunable gratings, where the typical response time is in the range 1-10 s or more.[55,56]

**3.3 Geometric and environmental control over the steering dynamics**

As anticipated in the previous section, the LCN light-driven actuation is mostly a two-step process, in which light is as first absorbed by the dye which dopes the liquid crystalline network, thus exciting its internal vibrational degrees of freedom. Subsequently the thermal energy diffuses in the lattice and warms up the whole structure, causing a total or partial lost of molecular orientation and driving a consequent reversible deformation. Once the excitation source is turned off, the grating temperature thermalizes with the thermal bath acquiring the initial shape. The observed time scale of the phenomenon reflects the thermal dynamics of an object of finite thermal capacitance subject to a heat flux, and in contact with a dissipative environment. A photonic designer can indeed exploit this effect by engineering the thermal environment of the beam steerer to optimize its temporal response: for instance, the grating can be surrounded by a high thermal conductivity liquid. To accurately describe the thermal diffusion in the LCN grating we employed a commercial software based on the finite element method (Comsol Multiphysics). We thus predict the time-dependent three dimensional thermal profiles of the structure upon actuation. The modeled grating structure is anchored to a glass substrate through LCN pedestals reproducing the fabricated structure by DLW with the same geometrical parameters. The temperature analysis has been retrieved for the excitation (control on) in form of a light beam that impinges on the grating for 1.4 ms, a shorter time that the one employed experimentally (5 ms) to reduce the calculation cost. It is then absorbed from the elastomeric structure following the Lambert-Beer law. After the stationary condition attainment, the light activation is removed (control off) and the structure thermally relaxes into the local environment, which we imposed to have a temperature of 300 °K. We tested the device response in air and water as reported in **Figure 4**. For the moment, no elastic deformation has been introduced into the simulation. Assuming a linear dependence of the deformation from the structure temperature, the mechanical LCN reshaping remains beyond the scope of our studies focused on the thermal dynamics. Instead, we accurately described the thermal response of the material by including the correct temperature dependency and spatial anisotropy of the thermal conductivity. We therefore evaluate the grating behavior in air and water and varying geometrical parameters as the number of pedestals and their height to better understand the thermal diffusion process. Consistently with the optical beam steering experiments described above, we first analyze the structure in air, to validate the simulative approach. In Figure 4a the modelled structure is reported in the inset and the steady-state temperature distribution inside and around the structure is plotted. In a low conductive medium as air (k = 0.03 W/m °K @ 300 °K), an important thermal exchange channel that links the grating to the glass substrate (k = 1 W/m °K) consists in the pillars even if is not the most effective. In fact, we verified that keeping constant the impinging intensity on the grating structure, the increase of the pedestal number from 3 to 7 corresponds to a local temperature increase of 16 °K, while taller pillars that push afar the structure from the glass substrate (comparison in between pillars of 1.5 µm and 10 µm in height), determine a temperature variation of more than 100 °K (Supporting Information). This is due to the good glass conduction and the large exchange surface. A more quantitative estimation of the grating distance effect from the glass substrate has been done increasing the pillars height, and controlling the impinging power to keep the stationary temperature constant of 400 °K. Positioning the substrate farther from the emitting grating, the heat dispersion decreases and a smaller activation power is necessary. This parameter obviously strongly affects also the temporal dynamics (Figure 4b). Once fixed the number of pedestals (3 as they correspond to the lower required power), the typical time-responses change significantly: response time becomes 3 times slower passing from a structure suspended on 1.5 µm to 15 µm. At the same time, also a reduction in the actuating intensity is recorded: only 4 µW/µm$^2$ are necessary to activate a grating suspended on 15 µm tall pedestals (13 µW/µm$^2$ for 1.5 µm tall pedestals). A complementary calculation about the effect of the number of pillar is reported in Supporting information.

To push the speed performances of our device, a liquid environment with higher conductivity as water (k=0.6 W/m°K @ 300 °K) has been introduced. Figure 4c shows the different temperature distribution into the grating surrounding area. Due to the water higher conductivity, the heat generated within the grating is efficiently transferred to the adjacent medium. Comparing the 3D temperature map in air (figure 4a) and water (figure 4c), it is evident how the heat distribution differs. In fact, in water a temperature of 400 °K is measured also on the glass surface, (instead of the measured 340 °K in air) and decreases into the substrate. In terms of device-relevant parameters, there is a sensible effect on the activation power: with respect to the previous case a value larger by nearly one order of magnitude (70 µW/µm$^2$) is needed to reach the stationary temperature of 400 °K. Meanwhile, and most importantly, the time-response in this environment is strongly shortened. Almost one order of magnitude can be gained by simply immersing the grating in water. Moreover, fine but interesting features can be observed by taking a careful look at the curves in Figure 4d. With respect to the air-embedded grating, a different temperature dependence is now present. The curves are now characterized by the sum of two exponentials with two different time constants. The double exponential rise and decay of the temperature profile can be attributed to the presence of two mechanisms: a fast dissipation into water, and a slower heat release through the polymer legs and the glass substrate. Also in this case, adjusting the grating separation from the glass and balancing the two dissipation contributes enables the realization of devices with different time-responses. To better depict the differences in air and water, the temperature map across the structure is shown in Figure 4e. It is clear how the conduction of water increases the heat flux out from the structure and the role of pedestals as thermal bridges vanishes. The computed temperature analysis retrieved for the excitation and relaxation process of the grating underlined the role of the environment on the structural properties and suggested the better configuration to be studied experimentally. The chosen structure is a diffractive grating structure suspended on a 3x3 pillar matrix of 1,5 µm and 5 µm height. Taller pedestals (10 µm and 15 µm) have a too large height/diameter ratio, resulting in quite unstable structures. Indeed, we partially succeeded in fabricating such structured devices (see Supporting information), suggesting that further chemical engineering of the material could lead to higher pedestal based beam steerers with an arbitrary geometry and able to respond to very little amounts of power. In the experimental data shown in **Figure 5**, the percent angular shift of the diffracted beams, due to the excitation, is reported. This value represents also the grating deformation time-dependence. Assuming that the grating shape changing is linearly dependent from the structure temperature, that rules the LCN phase transition and relative deformation, we can compare the experimental and calculated profile for the grating in air and in water.

For the 2D grating held up on 1,5 µm tall pedestals (Figure 5a), the maximum relative deviation of the diffracted order is around 10% for an excitation power of 20 mW. As the structure is operating in water, the performances show a relative deformation of 11% (under an actuation power of 220 mW). It is confirmed from this experimental result that the actuation power in water is one order of magnitude higher than in air due to the higher water thermal conductivity. It is also interesting to compare the experimental time-responses. Indeed, as retrieved from FEM calculations, the dynamics in water becomes faster both for the rise and the decay times, extrapolated from exponentially fitting the measured data. However, in this case, the time evolution is not characterized by a double exponential function, suggesting the dominating effect of water conductivity. The same characterization has been performed also for grating structures with 5 µm pedestals (Figure 5b) and the same trend can be found out. Comparing the two structures we retrieved the same expected behavior evaluated with the FEM calculations that show how the higher the pedestals the slower the time response is. However, the obtained results are one order of magnitude larger than the calculated one, likely because the real system dynamics is not only ruled by the time-response connected with heat diffusion but also from that of thermo-mechanical structure reshaping. It is interesting that despite of this, the simple thermal diffusion model can predict semi-quantitatively the effect of water, of the height of pedestal and the presence of a glass substrate.

**4. Conclusion and outlook**

We report on a new strategy to use light itself to control the optical response of photonic materials, using mechanical deformation as intermediate step. The materials consist of photo-responsive polymers structured by direct laser writing.

As example we prepared a two dimensional grating able to act as a fully optically driven beam steerer. The materials were characterized under various conditions revealing a very large optical response at sub-millisecond time scales. To better understand the potential of this technology, the time-response was compared with time-dependent thermal diffusion calculations. These calculations predicted that an even faster time-response can be obtained by embedding the materials in an environment with high thermal conductivity. We used these results to optimize our demonstrator device even further and to reduce the time-response by one more order of magnitude. The possibility to pattern optically-controlled micro beam steerers with a nanoprinting technique opens up the possibility to integrate them in various systems, including integrated photonic circuits. Another interesting development to explore is that of creating shapes exhibiting opto-mechanical bistability and/or memory type effects. While the optical grating serves as an example, the above described concepts can, in principle, be applied to a very broad range of optical materials with structural color and used to create large non-linear optical effects.

## 5. Experimental Section

*Liquid crystal cell preparation*: A polyimide layer (Nissan Chemical) is spin coated on cleaned coverslips and then rubbed by a velvet cloth. The cell height is fixed by 20 μm diameter glass sphere spacers. The LCN monomeric mixture is then melted on a hot plate at 70 °C, infiltrated in the glass cell, and allowed to slowly cool down keeping the homogeneous alignment for hours at room temperature.

*Direct Laser Writing*: The commercial DLW system (Nanoscribe) induces photopolymerization in photosensitive monomeric liquid mixtures using a 120 fs laser pulse at 780 nm that is focused through a 100x oil immersion objective with high numerical aperture (NA 1.4). The typical resolution, that depends on the material and the lithographic parameters, is around 100 nm.

*Experimental characterization*: For the grating diffraction characterization, we analyzed the far-field diffraction pattern of a red probe laser (633 nm He-Ne, ≈ 10 μw power) while exciting the grating with a CW green pump laser (532 nm diode laser, ≤ 20 mw power, filtered out while imaging the diffraction pattern). Both beams are focused to a spot approximately equal to the grating size.

*Temporal characterization*: To characterize the dynamical behavior of the beam, the diffraction pattern is not formed anymore on a screen, rather, it is directly imaged by a Fourier lens on the sensing element of a fast CCD camera (Photron FastCam SA4). In such a way, the signal level is much higher and very short integration times are sufficient to appropriately reconstruct the diffraction pattern, which can hence be recorded at a high frame rate (20000 fps). The control beam is chopped by a rotating wheel at 100 Hz; placing the wheel at the focus of a confocal lens pair allows to reach sharp rising and falling fronts of the excitation power trace down to about 0.05 ms. The excitation power is directly monitored by extracting a fraction of the control beam and sending it to a corner of the CCD element (the residual excitation beam exiting the sample is filtered out).

*Numerical calculation*: Modeling of the grating structure dynamics has been performed with a finite element method based commercial software, Comsol Multiphysics (RF+Heat Transfer modules). The grating has been designed with the geometrical parameters of the fabricated structure. The time dependent calculation describes the light that is then absorbed from the elastomeric structure following the Lambert-Beer law. After the stationary condition attainment, the light activation is removed (control off) and the structure thermally relaxes into the local environment (air or water), which we imposed to have a temperature of 300 °K.

**Supporting Information**

Supporting Information is available from the Wiley Online Library or from the author.

**Acknowledgements**

The research leading to these results has received funding from the European Research Council under the European Union's Seventh Framework Program (FP7/2007–2013)/ERC grant agreement n° 291349 on photonic micro robotics and Laserlab-Europe, H2020 EC-GA 654148; and from Ente Cassa di Risparmio di Firenze (2015/0781). We aknowledge Ce.M.E. (CNR Florence) for useful SEM images and Dott. Hao Zeng, Dott. Pietro Lombardi and Francesco Utel for their precious advices and discussions on the experimental measurements.

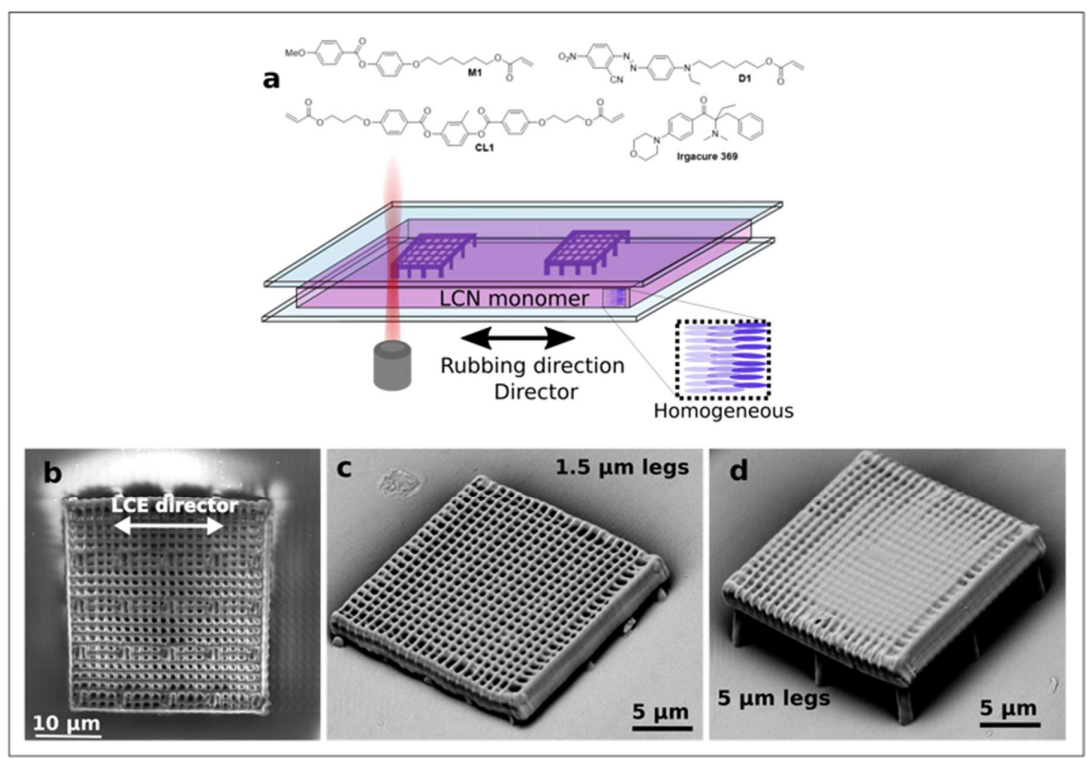

**Figure 1. Diffraction gratings.** a) Monomeric compounds (on the top) and fabrication scheme by Direct Laser Writing process inside the liquid crystalline cell (on the bottom). The resulting structure is reported in the scanning electron microscope images: respectively from the top (b) and with the substrate at 45° (c, d). Images (c) and (d) show gratings with two different leg height (1.5 μm and 5 μm). The number of legs is also different (5 legs per side in panels b-c, 3 legs per side in panel d).

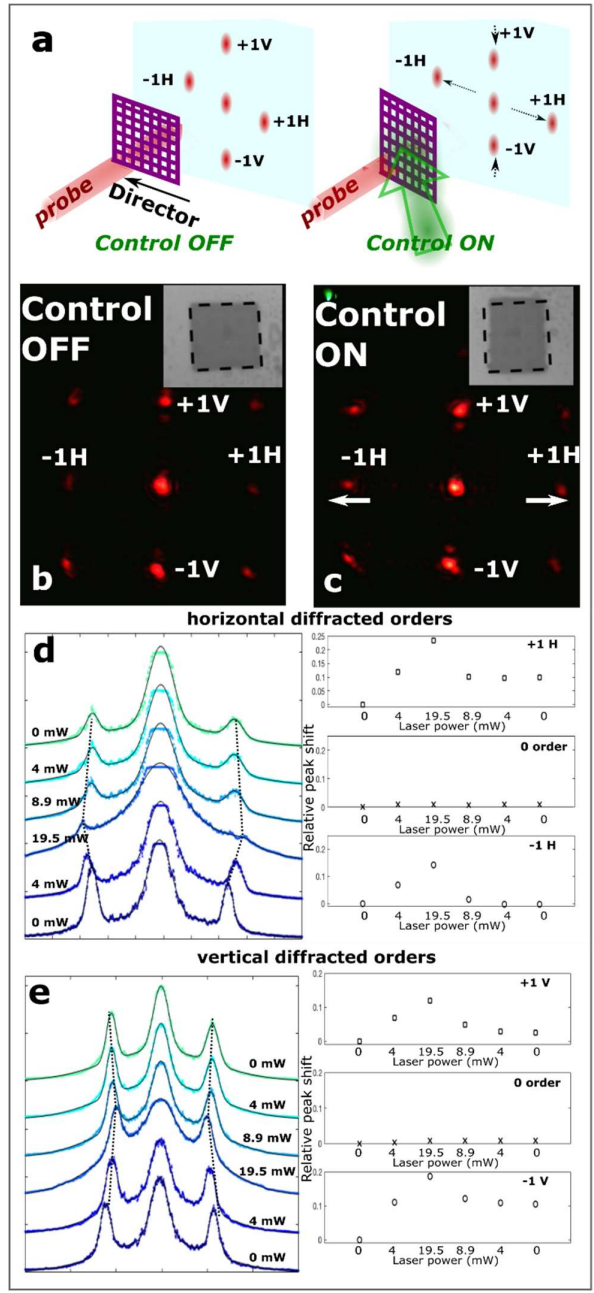

**Figure 2. Remotely controllable beam steering.** a) Operating scheme of the optical control over the diffraction angle. Diffraction pattern in absence of the control beam (b) and deformed diffraction pattern in presence of the control beam (c). The optical real image of the grating in the two configurations is reported in the insets. d) Horizontal diffracted peak profiles as a function of the control power. e) Vertical diffracted peak profiles. To the right of panels c) and d), the relative values of the peak shift (with respect to the rest position) are reported as a function of control power.

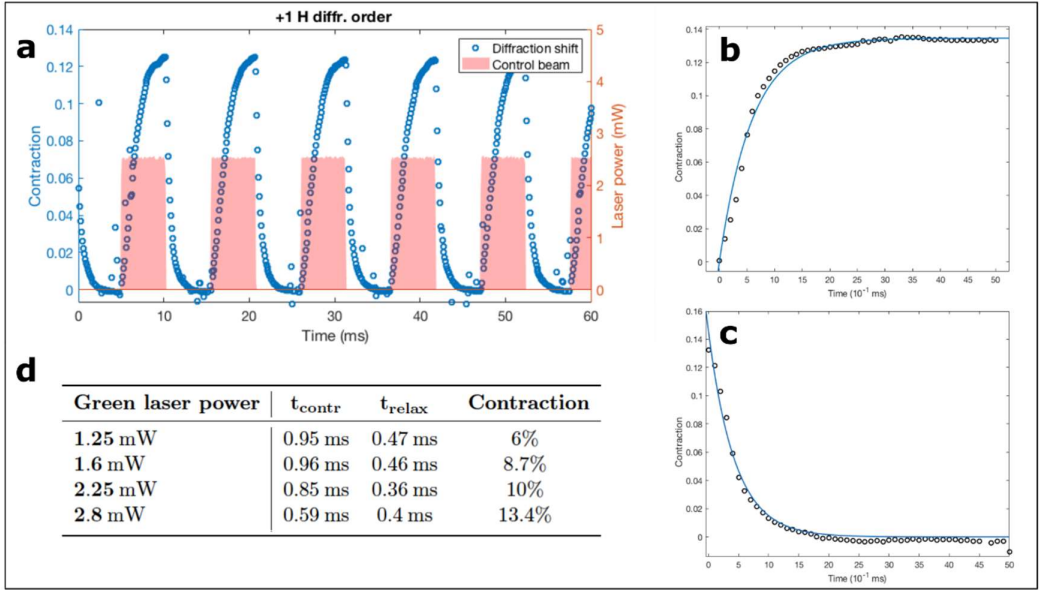

**Figure 3. Excitation and relaxation dynamics.** a) Temporal dynamics of the normalized deflection of the first diffraction order for an actuation power of 2.8 mW. Exponential rise and decay fitting for the switch-on process (b) and the switch-off (c). (d) Contraction and relaxation times (respectively $t_{contr}$ and $t_{relax}$) and maximum deflection of LCN diffraction gratings are extrapolated from the fitting procedures. To evaluate the power dependence, the diffraction measurements were performed by increasing the actuation power from 1.25 mW up to 2.8 mW.

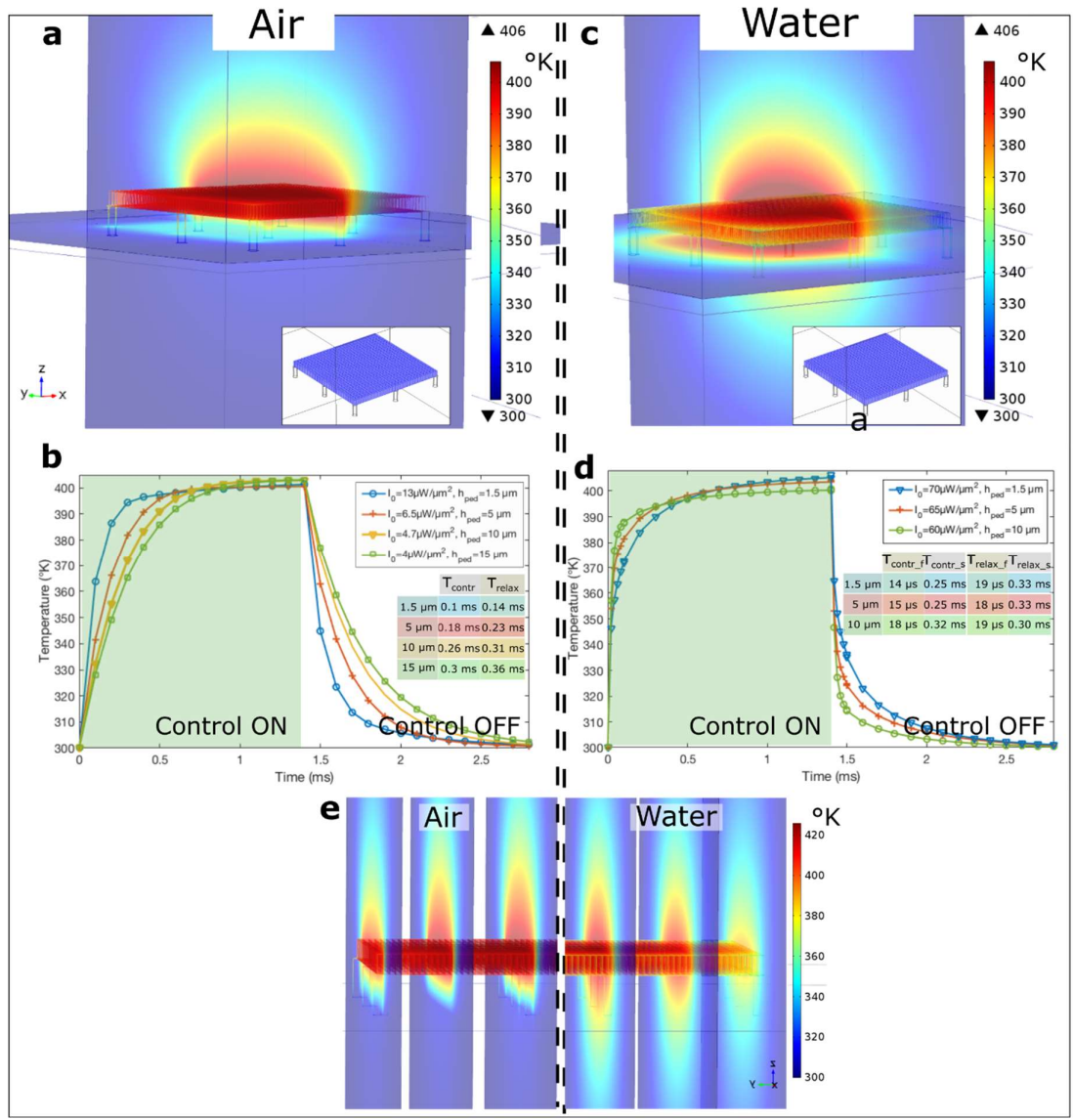

**Figure 4. Thermal calculation results for the grating immersed in air and in water.** The results of the FEM calculation are summarized on the left for the structure in air, and on the right in water. A 3D map of the temperature profile is reported once the grating structure (suspended on 3x3 of 5 µm tall pedestals) reaches a stationary temperature after illumination in air (a) and water (c). The central row describes the grating average temperature time evolution. Contraction and relaxation times (respectively $t_{contr}$ and $t_{relax}$) are evaluated by fitting the exponential profiles. The results are reported for the grating structure in air (b) and water (d) varying the height of the pillars on which is suspended. In water, a double exponential curve has been used to fit the data underlining the presence of a fast contribute ($t_{contr\_f}/t_{dec\_f}$) and a slower one ($t_{contr\_s}/t_{dec\_s}$). To compare the thermalization process in the two different fluids, temperature profiles (e) are reported.

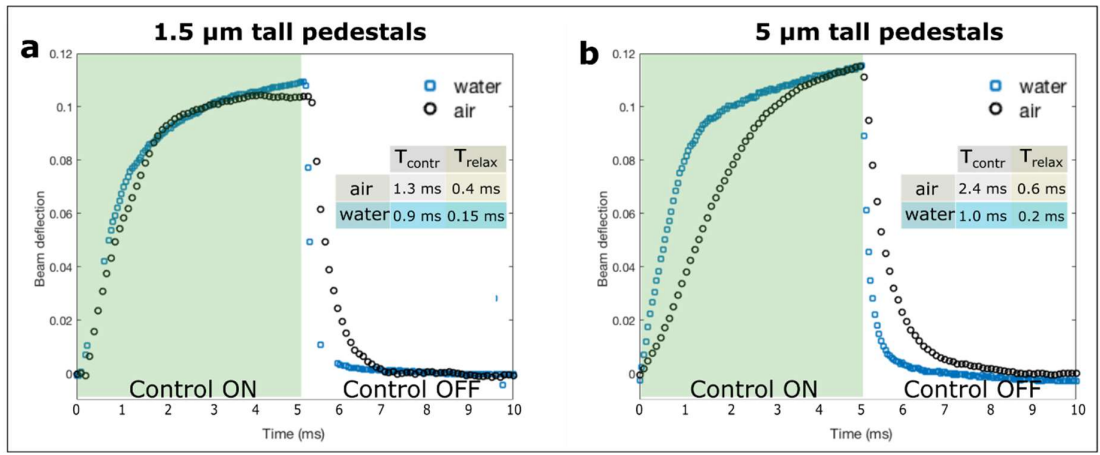

**Figure 5. Experimental results: air and water dynamics.** The contraction time dependence is reported for a soft grating suspended on 3x3 pedestal whose height is a) 1.5 µm and b) 5 µm both in air (black circles) and in water (blue squares). In the tables, the time constant values retrieved from single exponential fitting curves are reported.

Supporting Information for the article
# Structured Optical Materials Controlled by Light

By Sara Nocentini, Daniele Martella, Camilla Parmeggiani, Simone Zanotto, and Diederik S. Wiersma

**Figure S1.** Scanning Electron Microscope (SEM) images of unsuccessful fabrication of grating beam steerers. Figure a) shows 10 µm tall pedestal grating structure. b) and c) report fallen gratings on the glass substrate. In this case, due to the too large height/diameter ratio of the pillars, the beam steerer body collapses and remaining adhesive to the glass surface, it loses its photoresponsive shape-change property.

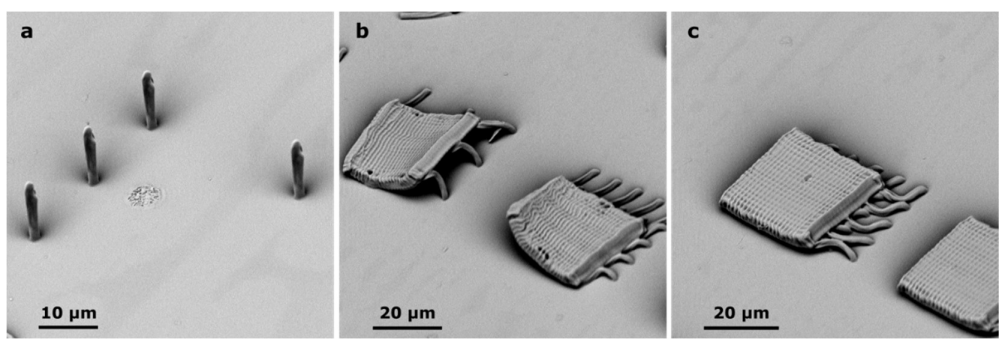

**Figure S2.** Temporal dynamics of the grating structure calculated with a commercial FEM based software (Comsol Multiphysics). Varying the impinging light intensity and the number of pedestals holding the grating structure, the time response is evaluated. The height of the pillars has been fixed at 5 µm. As the graph illustrates, the number of pedestals does not affect considerably the time response while it increases the thermal exchange with the substrate requiring a higher energy to reach an average temperature of 400 °K of the grating.

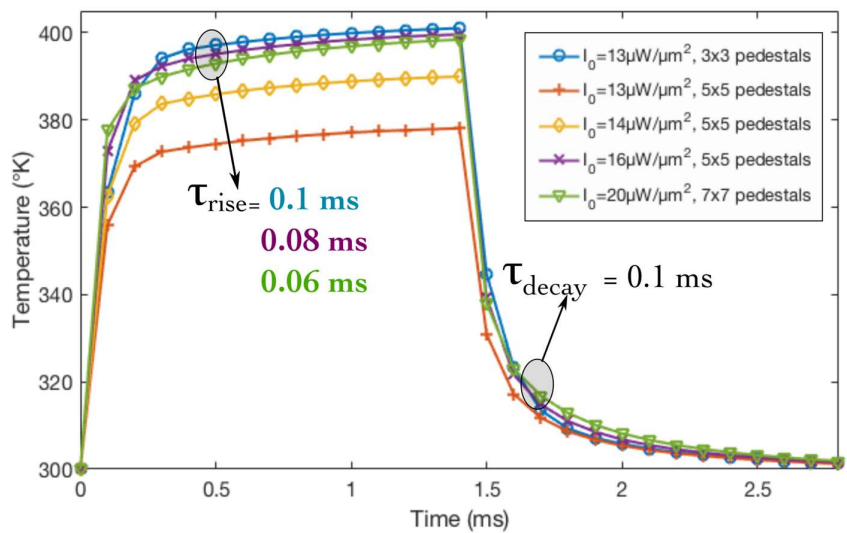

**Figure S3.** FEM calculation results of the temporal actuation dynamics. Keeping the imping power value constant and varying the number of pillars and their height demonstrates as the more important contribution is related to the distance of the grating from the substrate.

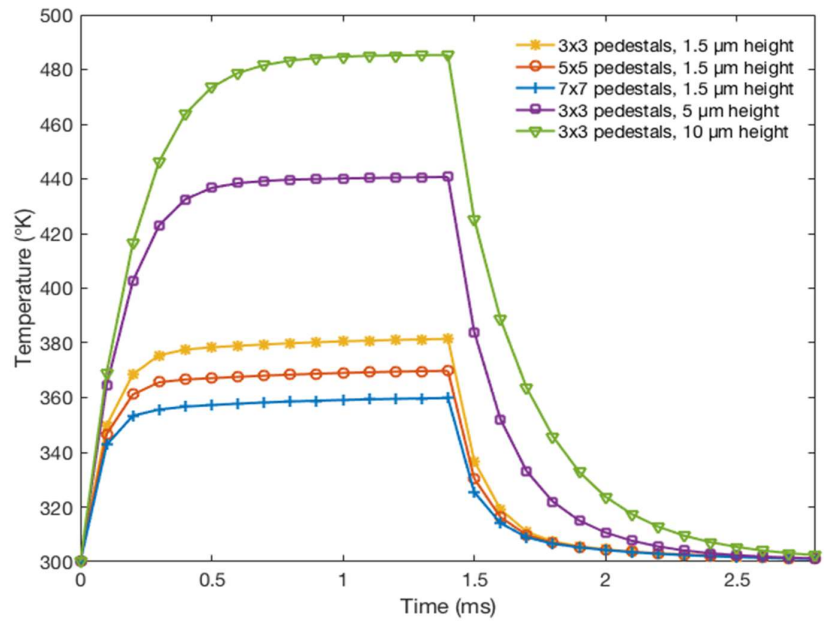